\documentclass{elsarticle}

\usepackage{amssymb, amsmath}
\usepackage{graphics, array, url, hyperref, epsfig, amsmath,amssymb}
\usepackage{graphicx}
\usepackage{subfigure}
\usepackage{CJK}
\usepackage{latexsym,amsfonts}
\usepackage{tabularx}
\usepackage{epstopdf}
\usepackage{fancybox}
\usepackage{multirow}
\usepackage{array}
\usepackage[inline,draft,nomargin]{fixme}

\begin{document}

\begin{frontmatter}



\title{An Efficient Linkable Group Signature for Payer Tracing in Anonymous Cryptocurrencies}

\author[snnu]{Lingyue Zhang}
\author[snnu]{Huilin Li}
\author[uow]{Yannan Li}
\author[snnu]{Yanqi Zhao}
\author[snnu]{Yong Yu\corref{corresponding author}}
\cortext[correspondingauthor]{Corresponding author}
\ead{yuyong@snnu.edu.cn}





\address[snnu]{School of Computer Science, Shaanxi Normal University, Xi'an, 710062, China.}
\address[uow]{School of Computing and Information Technology, University of Wollongong, Wollongong, NSW 2522, Australia.}

\begin{abstract}
Cryptocurrencies, led by bitcoin launched in 2009, have obtained wide attention due to the emerging Blockchain in recent years. Anonymous cryptocurrencies are highly essential since users want to preserve their privacy when conducting transactions. However, some users might misbehave with the cover of anonymity such as rampant trafficking and extortion. Thus, it is important to balance anonymity and accountability of anonymous cryptocurrencies. In this paper, we solve this issue by proposing a linkable group signature (LGS) for signing cryptocurrency transactions, which can be used to trace a payer's identity in consortium blockchain based anonymous cryptocurrencies, in case the payer tries illegal activities. A payer keeps anonymous if he/she behaves honestly. We prove that the proposed scheme achieves full-anonymity, full-traceability and linkability in the random oracle. Implementation of the proposed LGS scheme demonstrates its high efficiency thus, can be adopted in anonymous cryptocurrencies in reality.
\end{abstract}

\end{frontmatter}

\section{Introduction}\label{sec1}
Cryptocurrency is a medium of exchange by applying cryptographic techniques to enable secure transactions. A significants difference between traditional e-cash system and cryptocurrencies is the former need a central third party while the latter do not. In cryptocurrencies, transactions are recorded in through a distributed public ledger called Blockchain\cite{Nakamoto}, which has attracted extensive attention from academia and industry. Several papers \cite{Du1,Du2,Du3,Du4,Du5,Du6,Du7,Du8,Du9} have studied related security issues.

Anonymity, which protects users' privacy in cryptocurrency transactions, is one of the most prominent characteristics of cryptocurrencies. 
In the meantime, it may also lead to some new threats such as rampant trafficking, extortion, smugglers' business, tax evasion and money laundering. Worse still, it is hard to trace the criminals with the cover of anonymity. Unfortunately, criminal activities using cryptocurrencies have emerged in recent years. An example 
is Silk Road, a black market for illegal drugs and illicit goods. It trades by using Bitcoin\cite{Nakamoto} for payment and hides service with Tor \footnote{https://www.onion-router.net/}.
Another example is NotPetya \footnote{https://zh.wikipedia.org/wiki/NotPetya} ransomware, which was used to launch a global cyber attack and hacked over 80 companies in a host of countries led by Ukraine. NotPetya extracts passwords from local files by encrypting the entire hard disk, and then victims were asked to pay \$300 in cryptocurrency to unlock their computers. Maersk, the world's largest container ship and supplier, has caused a loss about \$200-\$300 million due to the business interruption caused by NotPetya. It is sad that the wrecker has not been catched till now. Hence, anonymous cryptocurrencies must be regulated to prevent the abuse of anonymity.

Traceability is an effective way for supervision. Most cryptocurrencies such as Bitcoin record transaction history on a pubic Blockchain, where anyone can join, access and leave anytime. 
This kind of cryptocurrencies usually protect users' privacy with a pseudonyms mechanism, in which
a user is identified by the hash of his/her public key and a user could generate a multitude of public keys as their deterministic wallets\cite{maxwell}-\cite{electrum}. What can be traced of a suspicious user is the transaction addresses or wallets instead of his/her real-world identities under this pseudonyms mechanism. Actually, the traced addresses are the hashed public key of the user, i.e., random bit-strings. Only tracing addresses or wallets are not enough in some real-world applications. For example, Monero\cite{noether}, another popular cryptocurrency, takes advantage of one-time address mechanism to protect payee' privacy. It makes no sense to trace this one-time address when the payee becomes a malicious payer since this address will not be used any longer. In public blockchain based anonymous cryptocurrency systems, it is hard to trace the real identity of a user due to the employment of pseudonyms mechanism. Consortium blockchain based anonymous cryptocurrencies provide a potential solution to this problem.


Group signature\cite{bon}-\cite{camen}, which achieves traceability through a group manager, 
can be applied to consortium blockchain based anonymous cryptocurrencies tracing. Users register
with the group manager as a new group member, then he can sign transactions represent the whole group without exposing identity, where group manager can trace the suspectable users in a anonymity cryptocurrency trading.
Due to the anonymity provided by group signatures, naturally, we hold the view that how to determine two transactions are generated by the same signer instantly and effectively.



\textbf{{Related Work.}}
Koshy et al. \cite{koshy} constructed a mapping from Bitcoin accounts to IP addresses according to examine the real-time transactions to find the account owners. Reid et al. \cite{reid} acquired a sea of transactions from public Blockchain and try to model the Bitcoin flow, they attempt to link multiple addresses to the identical user by analyzing their topology structure and addresses reusing. Kumar et al. \cite{kumar}introduced a statistical analysis regarding Monero transactions tracing. The success rate of traceability is limited by the number of users in the anonymity trading set. In 2016, Danezis et al. \cite{danezis} proposed RSCoin, a cryptocurrency based on central banks who completely involved in controlling monetary policy, had been issued by the bank of England. In 2017, Cecchetii el at. \cite{cec} presented a confidential distributed ledger named $Solidus$, where all users' accounts are maintained by banks and transactions are executed between diverse banks. Users whether payers or payees can not be openly traced even their pseudonyms except for the bank that owners their accounts. In addition, Garman et al. \cite{garman} introduced a tracing mechanisms supporting optional users tracing and cryptocurrency tracing by utilizing cryptographic tools. Nevertheless, the solutions aforementioned can only trace to user's accounts instead of their real-world identities.


\textbf{{Our Contributions.}} The contributions of this paper are listed as follows.
\begin{enumerate}
  \item We suggest to use linkable group signatures to realize payers' real identity tracing in consortium blockchain based anonymous cryptocurrency systems. This approach provides a tradeoff between anonymity and accountability in anonymous cryptocurrencies.
  \item We propose a concrete construction of linkable group signature based on the group signature due to Boneh and Boyen \cite{boneh}. The proposed scheme makes use of linear encryption to help the group manager to trace a group member's identity, and generates a group signature by generating a zero-knowledge proof of knowledge (ZKPK) of a triple tuple of VR-SDH. If a group member signs the same message twice, the two signatures can be publicly linked, which can be used for double-spending detection in anonymous cryptocurrencies.
  \item We prove the security of our linkable group signature scheme including full-anonymity, full-traceability and linkability in the random oracle model.
  \item We implement the proposed linkable group signature on the desktop, which shows its practicability in reality.
\end{enumerate}

\textbf{Organization.} The rest of this paper is organized as follows. Some preliminaries of our scheme are prepared in Section \ref{sec2}. We describe our system model and security requirements in Section \ref{sec3}. The details of our scheme is provided in Section \ref{sec4}, then it followed the security analysis and performance evaluation are given in Section \ref{sec5} and Section \ref{sec6} respectively. Finally, we conclude our paper in Section \ref{sec7}.

\section{Preliminaries}\label{sec2}
\subsection{Bilinear Groups}
The security of our scheme rely on the assumption of bilinear group. We say $(G_{1},G_{2})$ be a bilinear group if following properties holds: $G_{1}$,$G_{2}$ denote two multiplicative cyclic groups of prime order $p$, satisfied $G_{1}=<g_{1}>$ and $G_{2}=<g_{2}>$. A computable isomorphism $\varphi$ is constructed from $G_{1}$ to $G_{2}$ such that $g_{1}=\varphi(g_{2})$. $\hat{e}$ be a bilinear map \cite{boneh1} $\hat{e}:G_{1}\times G_{2}\longrightarrow G_{T}$ such that:

\emph{Bilinearity.} $\hat{e}(h_{0}^{a},h_{1}^{b})=\hat{e}(h_{0},h_{1})^{ab}$ holds for all $h_{0}\in G_{1}$, $h_{1}\in G_{2}$ and $a,b\in Z_{p}$.

\emph{Non-degenerate.} $\hat{e}$ does not map $h_{0}, h_{1}$ to the identity $1_{G_{T}}$ such that $\hat{e}(h_{0},h_{1})\neq 1_{G_{T}}$, where $1_{G_{T}}$ is the identity of group $G_{T}$. \\
It is efficient to compute $\hat{e}(h_{0},h_{1})$, isomorphism $\varphi$ and $G_{1}, G_{1}$ as aforementioned, whether $G_{1}=G_{2}$ or $G_{1}\neq G_{2}$.

\subsection{Complexity assumption}
\noindent \textbf{\emph{(q-Strong Diffie-Hellman Problem.)}} Boneh and Boyen \cite{boneh2} defined q-Strong Diffie-Hellman Problem in the bilinear group pair $(G_{1},G_{2})$ stating that given a $(q+2)$-tuple $(g_{1},g_{2},g_{2}^{\gamma},g_{2}^{(\gamma^{2})},... ,g_{2}^{(\gamma^{q})})$ for random $\gamma\leftarrow Z_{p}$,
it is impracticable to output a pair $(g_{1}^{1/(\gamma+x)},x)$ where $x\leftarrow Z_{p}$. \\
The advantage of adversary $\mathcal{A}$ to settle $q$-SDH problem defined as follow:
$$Adv_{\mathcal{A}}^{SDH}=Pr\Big[\mathcal{A}(g_{1},g_{2},g_{2}^{\gamma},g_{2}^{(\gamma^{2})},... ,g_{2}^{(\gamma^{q})})=(g_{1}^{\frac{1}{(\gamma+x)}},x)\Big]$$

\textbf{Definition 1.}(q-Strong Diffie-Hellman in $(G_{1},G_{2})$)
\emph{We say that $(q,t,\epsilon)-$SDH assumption holds in $(G_{1},G_{2})$ if no $t$-time adversary solves SDH with advantage $Adv_{\mathcal{A}}^{SDH}$ at least $\epsilon$ in $(G_{1},G_{2})$.}

\noindent\textbf{\emph{(q-Variant Strong Diffie-Hellman Problem.)}} Fuchsbsuer et al.\cite{fu} introduced the hardness of SDH implies that the following problem is intractable: given $g_{1},g_{2},h,g_{2}^{\gamma}$ and $q-1$ distinct triples $(x_{i},y_{i},(g_{1}\cdot h^{-y_{i}})^{\frac{1}{\gamma+x_{i}}})$, output a fresh triple $(x,y,(g_{1}\cdot h^{-y})^{\frac{1}{\gamma+x}})$.  \\
The advantage of adversary $\mathcal{A}$ to settle $q$-VR-SDH problem defined as follow:
$$Adv_{\mathcal{A}}^{VR-SDH}=Pr\Big[\mathcal{A}((x_{1},y_{1},(g_{1}\cdot h^{-y_{1}})^{\frac{1}{\gamma+x_{1}}}),... , (x_{q-1},y_{q-1},(g_{1}\cdot h^{-y_{q-1}})^{\frac{1}{\gamma+x_{q-1}}})$$
\hspace{2.9cm}$=(x,y,(g_{1}\cdot h^{-y})^{\frac{1}{\gamma+x}})\Big]$

\textbf{Definition 2.}(Variant q-Strong Diffie-Hellman in $(G_{1},G_{2})$)
\emph{We say that $(q,t,\epsilon)-$VR-SDH assumption holds in $(G_{1},G_{2})$ if no $t$-time adversary solves VR-SDH with advantage $Adv_{\mathcal{A}}^{VR-SDH}$ at least $\epsilon$ in $(G_{1},G_{2})$.}

\noindent\textbf{\emph{(Decision Linear Problem)}} Boneh and Boyen \cite{boneh} proposed it is infeasible to solve the problem as follows: given $v_{1},v_{2},u,v_{1}^{a},v_{2}^{b},u^{c}\in G_{1}$, output 1 if $a+b=c$, otherwise 0. \\
The advantage of adversary $\mathcal{A}$ to distinguish the DL Problem is defined as follows:
$$Adv_{\mathcal{A}}^{Linear}=Pr\Big[\mathcal{A}(v_{1},v_{2},u,v_{1}^{a},v_{2}^{b},u^{a+b})=1:v_{1},v_{2},u\in G_{1},a,b\in Z_{p} \big]$$
\hspace{3.2cm}$-\Big[\mathcal{A}(v_{1},v_{2},u,v_{1}^{a},v_{2}^{b},\eta)=1:v_{1},v_{2},u,\eta\in G_{1},a,b\in Z_{p} \big]$

\textbf{Definition 3.}(Decision Linear in $G_{1}$)
\emph{We say that Decision Linear assumption holds in $G_{1}$ if no $t$-time adversary $Adv_{\mathcal{A}}^{Linear}$ decides DL problem with advantage at least $\epsilon$ in $G_{1}$.}

\subsection{Signature of Knowledge}
CS\cite{cam} advocated the primitive of the signature of knowledge for the first time, which is a transformation from a interactive proof system to a non-interactive proof system by letting the challenge equals the hash value of the commitment concatenate the message to be signed. Take $\sum-$protocol as an example, $\sum=(Setup, Sign, Verify)$ be a signature of knowledge for a NP-relation $\mathcal{R}$, where language $L=\{h:\exists  \omega, s.t.(\omega,h)\in \mathcal{R}\}$. the specific algorithm is showed as follows:

\begin{itemize}
\item \emph{$Setup(1^{\lambda})$:} This algorithm takes security parameter $\lambda$ as inputs, and output a public parameter $pp$.

\item \emph{$Sign(m,\omega,h)$:} This algorithm takes message $m$ and a pair $(\omega,h)\in \mathcal{R}$ as inputs, and output a witness $\Pi$ of signature of knowledge.

\item \emph{$Verify(m,h,\Pi)$:} This algorithm takes message $m$, a statement $h$ and a signature of knowledge $\Pi$ as inputs, and output a bit $b\in \{0,1\}$.
\end{itemize}

Chase et.al\cite{chase} defined the formal security of SoK, which is called $SimExt-secure$. A $SimExt-secure$
SoK of a witness $\omega$ for language $L$ consists of the following properties:

\textbf{Correctness.} For any relation satisfied $(\omega,h)\in \mathcal{R}$ with message $m$, there exists a negligible function $f$ such that
$$Pr\Big[Ver(m,h,\Pi)=1: pp\leftarrow Setup(1^{\lambda}), \Pi\leftarrow Sign(m,\omega, h)\Big]\geq 1-f(\lambda)$$

\textbf{Simulatability.} There exits a polynomial time simulator s.t it is infeasible to distinguish the transcription of simulator which denotes as $Sim=(Simsetup,Simsign)$ and the real protocol transcription for any PPT adversaries $\mathcal{A}$, for a negligible function $f$ such that
$$ \Big\arrowvert Pr[(pp,\theta)\leftarrow Simset(1^{\lambda}); b\leftarrow \mathcal{A}^{Sim}(pp)]-$$
$$ Pr[pp\leftarrow Gen(1^{\lambda}); b\leftarrow \mathcal{A}^{Sign}(pp)]\leq f(\lambda)\Big\arrowvert$$ \\
Where $\theta$ denotes an traditional trapdoor, in which the simulator to simulate a signature under the situation without the witness $\omega$.

\textbf{Extraction.} There exists another polynomial time extractor called $Ext$, which can extract the witness $\omega$ when he knows a trapdoor $\theta$ and a pair transcription of simulator. For a negligible function $f$ such that
$$Pr\big[(\omega,h)\in \mathcal{R}\vee (m,h)\in Q_{S} \vee Ver(m,h,\Pi)=0:(pp,\theta)\leftarrow Simsetup(1^{\lambda}),$$
$$(m,h,\Pi)\leftarrow \mathcal{A}^{Sim}(pp), \omega\leftarrow (Ext(pp,\theta,m,h,\Pi))\big]\leq f(\lambda)$$\\
Where $Q_{S}$ is the list of $Simsign$ Oracle query, which is enumerated the successful queries made by $\mathcal{A}$.

\subsection{Linear Encryption}
In our scheme, we utilize the linear encryption scheme, which is the extension of ElGamal encryption scheme based on decision linear problem in Section \ref{sec2}. The details of the algorithm are as follows.

\begin{itemize}
\item \emph{$KeyGen(\lambda)$:} This algorithm takes security parameter $\lambda$ as inputs, it randomly selects generators $v_{1},v_{2},u \in G_{1}$ and sets $k_{1},k_{2} \in Z_{p}$ as private key satisfied $v_{1}^{k_{1}}=v_{2}^{k_{2}}=u$, it outputs the key pair $(p=(v_{1},v_{2},u),s=(k_{1},k_{2}))$.

\item \emph{$Enc(m,p)$:} This algorithm takes message $m$, public key $p$ as inputs, it chooses random value $x,y \in Z_{p}$ and encrypt $m$ as $c=(v_{1}^{x},v_{2}^{y},m\cdot u^{x+y})$. it outputs the ciphertext of $m$ be $c$.

\item \emph{$Dec(c,s)$:} This algorithm takes message $m$, ciphertext $c$ and the private key $s$ as inputs, $m=c_{3}/(c_{1}^{k_{1}}\cdot c_{2}^{k_{2}})$ is computed by the one who knows the secrete key $(k_{1},k_{2})$. it outputs $m$.
 \end{itemize}

It is widely believed that ElGamal encryption against a chosen-plaintext attack. As an extension of ElGamal encryption,  linear encryption provides the same level of security as ElGamal emcryption under decision linear assumption.

\section{Models and Requirements}\label{sec3}
\subsection{System Model}
There are three entities in our scheme. Users register with the registration authority(RA) to be a legal member in a group as shown in Figure \ref{fig2}. Users could be vicious, which means that he may try to fraud RA to obtain a real certificate and forges a signature. RA, who takes charge of the private key of enrollment, can award
a certificate to a honest users and he is defined as honest, it means that he follows the protocol to perform the tasks allotted to him, meanwhile he possesses a registration list, which storages the identity of group members. There exists a supervision authority(SA), who is in charge of the private of tracing. 
he will regulate the behavior of illegal users by the way to trace the specific users' identity. Two signatures signed by the same user could be publicly linked by all members in a group.
  \begin{figure}[h]
  \begin{center}
  \includegraphics[width=0.8\textwidth]{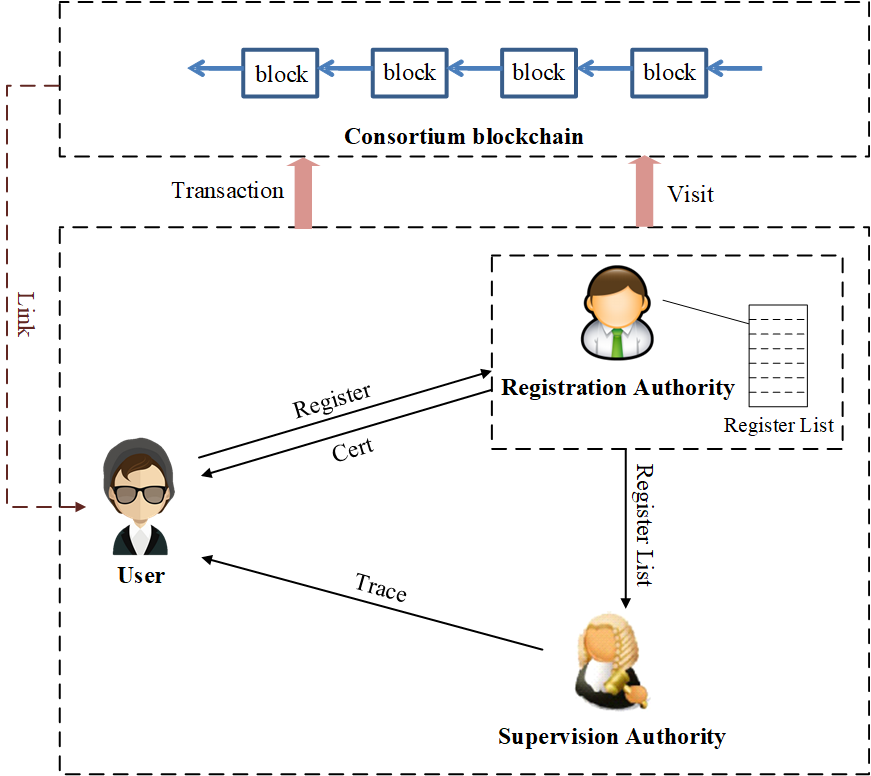}\\
  \caption{System model}\label{fig2}
  \end{center}
  \end{figure}





\textbf{Definition 4.} (Linkable group signature). A linkable group signature includes a tuple of  polynomial time algorithms ($Setup$, $Join$, $Sign$, $Verify$, $Link$, $ Trace$,) such that:
\begin{itemize}
\item \emph{$(GPK)\leftarrow Setup(n)$}: This is a probabilistic algorithm that inputs a security parameter $n$ and outputs the group public parameters $GPK$.
\item \emph{$(Cert,GSK_{U})\leftarrow Join(GPK,SK_{R},PK_{U})$}: This is a protocol between RA and users, it takes $GPK$, the private $SK_{R}$ of RA and user's public key $PK_{U}$ as inputs, outputs the corresponding membership certificate $Cert$, and membership private key $GSK_{U}$ of user.
\item \emph{$(\sigma)\leftarrow Sign(GPK,GSK_{U},m,amount)$}: This is a probabilistic algorithm that inputs $GPK$, membership private key $GSK_{U}$, message $m$ and an amount $amount\in \{0,1\}^{\ast}$ outputs a signature $\sigma$.
\item \emph{$(1/0)\leftarrow Verify(GPK,m,\sigma,amount)$}: The algorithm inputs $GPK$, message $m$, a alleged signature $\sigma$, an amount $amount\in \{0,1\}^{\ast}$ and outputs a bit $b\in \{0,1\}$.
\item \emph{$(1/0)\leftarrow Link(GPK,(m,\sigma), (m',\sigma'))$}: The algorithm takes the public parameter $GPK$, two tuples of signature $(m,\sigma)$, $(m',\sigma')$ as inputs and outputs a bit $b\in \{0,1\}$. \item \emph{$(PK_{U},Cert)\leftarrow Trace(GPK,SK_{S},m,\sigma)$}: The algorithm takes $GPK$, message $m$, a valid signatures $\sigma$ as inputs and outputs the users' identity $PK_{U}$ and certificate $Cert$. \end{itemize}

\subsection{Security Requirements}
There are diversiform security requirements of group signature has been proposed. Among all of them, two crucial properties be summarized by Bellare et.al.\cite{bel} in 2003 and they showed all other requirements are implied by them, which is full-anonymity and full-traceability. In 2018, Wu et.al.\cite{bel} introduced the security requirements of the linkable group signature, which is correctness, full-anonymity,
full-traceabitity, and linkability. We follow this formal definition, but, their scheme achieves conditional linkability while our scheme could reach public linkability.

\textbf{Correctness}: The correctness ensures that the signature generated by honest users always be accepted, two signatures produced with the same secret key could be linked correctly all the time, and any valid signature can be traced to the actual signer invariably with SA's tracing key.

\textbf{Full-Anonymity}: Given two signatures and a signer either one of them, no one could determine which of
two signatures was produced by the known signer with advantage over one-half except SA. Boneh et.\cite{boneh} showed a relax definition named CPA-full-anonymity, which cancels accessing to the tracing oracle. They also explained this requirement is enough.

\textbf{Full-Traceability}: Full-traceability is stronger than traceability, it also could be viewed as a strong form of collusion-resistance. Specifically, an adversary created a signature according to the collusion of other group members even holds the SA's trace key cannot be traced to one of the actual signer by SA with negligible probability.

\textbf{Linkability}: An adversary generated two signatures with the same secret key for the same amount without being linked by other group members with negligible probability.



\section{The Linkable Group Signature Construction}\label{sec4}
\subsection{Overview}
The system works as follows. First, the Registration Authority and the Supervision Authority initialize the system parameters. Then, users register with the RA as new group member. When users register on the RA, he is awarded a certificate, in which is the part of his private key in a group. As a result, only segmental users' private key is known by RA, and the rest is only known by himself. Next, users produce a signature on one message with a linkable group signature. Two signatures for the same amount can be publicly linked, while only the SA has right to trace who is the misbehaving users. Finally,
SA reals users' identity with his tracing secret key, and gets the index $i$ in a group according to the registration list given by the RA. 

\subsection{The details of protocol}\label{subsec:protocol}

\begin{itemize}
\item \emph{\textbf{$Setup$}:}

 1.$Init(Common\,parameter):$ Let $n$ be a security parameter, and $\hat{e}$ be a bilinear pairing. $(G_{1},G_{2})$ denotes a bilinear group pair with computable isomorphism $\varphi$. Assume SDH problem is hard on $(G_{1},G_{2})$ meanwhile decision linear problem is intractable on $G_{1}$.  Define $H_{0}:\{0,1\}^{\mathcal{*}}\rightarrow G_{1}$, $H_{1}:\{0,1\}^{\mathcal{*}}\rightarrow Z_{p}$, denote  collision-resistant hash. Randomly choose a generator $g_{2}$ in $G_{2}$, choose generator $g_{1},h,u$ in $G_{1}$ such that $g_{2}=\varphi(g_{1}).$

 2.$RA\,Setup:$ The RA randomly selects $\gamma\ \in_{R}\,Z_{p}$, and sets $\omega=g_{2}^\gamma$. $\gamma$ is the private key of RA. So that $PK_{R}=\omega,$ $SK_{R}=\gamma.$

 3.$SA\,Setup:$ The SA chooses a generator $v_{1},v_{2}\in_{R}\,G_{1}$. Select $k_{1},k_{2}\,\in_{R}\, Z_{p}$ such that ${v_{1}}^{k_{1}}={v_{2}}^{k_{2}}=u$. It has $PK_{S}=u,$ $SK_{S}=(k_{1},k_{2}).$

 Group public parameter be \\
$$GPK=(G_{1},G_{2},\hat{e},p,n,g_{1},g_{2},h, u, v_{1}, v_{2},\omega,H_{0}, H_{1} ).$$

\item \emph{\textbf{$Join$}:} User who expect to join the group must register RA at first, and gets corresponding certificate as a new group member. The concrete protocol as follows:

    1.User randomly selects $y_{i}\in \ Z^{\ast}_{p}$ sends $Y=h^{y_{i}}$ to RA, in the meantime, he shows the knowledge of the representation of $Y$ to bases $h$:
 $$ PK\{(y_{i}):Y=h^{y_{i}}\}$$

    2.RA randomly select $x_{i}\in_{R}\,Z_{p}^\ast$, sets $A_{i}={(g_{1} Y^{-1})}^{\frac{1}{\gamma+x_{i}}}$ with his private key $SK_{R}=\gamma$ and sends $(A_{i},x_{i})$ to user.

    3.User checks whether $\hat{e} \ (A_{i},\omega{g_{2}^{x_{i}}})=\hat{e} \ (g_{1}{h^{-y_{i}},g_{2}})$. If true, he agrees $(A_{i},x_{i})$ as his $Cert$, sends $Cert$ and $Y=h^{y_{i}}$ to RA, storages tuple $(A_{i},x_{i},y_{i})$ be his private-key $SK_{U}[i]$.

    4.RA maintains a registration list and $(Cert,Y)$ is added.

\item \emph{\textbf{$Sign$}:} In order to sign message $m\in\{0,1\}^{\mathcal{*}}$ with a group of $\{A_{1},A_{2},\ldots,A_{n}\}$, where $A_{i}$ denotes the $i^{th}$ users' certificate. Using users' private key $SK_{U}[i]=(A_{i},x_{i},y_{i})$ for the specified $amount\in\{0,1\}^{\mathcal{*}}$, compute $u_{0}=H_{0}(amount)$.

    Randomly choosing $\alpha,\beta$, and sets
    \begin{center} $l_{1}=v_{1}^{\alpha}$, \quad $l_{2}=v_{2}^{\beta}$, \quad $l_{3}=A_{i}\cdot u ^{\alpha+\beta}$, \quad  $l_{4}=u_{0}^{x_{i}}$

   $\delta_{1}=x_{i}\cdot \alpha$, \quad  $\delta_{2}=x_{i}\cdot{\beta}$ \end{center}

    Denote $1_{G_{1}}$ is the identity element of $G_{1}$. Then the signer executes the non-interactive zero-knowledge proof-of-knowledge $\Pi$ on $m$ as follows:
\end{itemize}

\begin{center}
$SoK\left\{
\left(
\setlength{\arraycolsep}{0.5pt}
\begin{array}{ll}
\alpha,&\beta, \\
x_{i},&y_{i},\\
\delta_{1},&\delta_{2}
\end{array}
\right)
:
\begin{array}{cc}
                     &\ l_{1} = v_{1}^{\alpha} \\
              \wedge &\ l_{2} = v_{2}^{\beta} \\
              \wedge &\ 1_{G_{1}} = l_{1}^{x_{i}}\cdot v_{1}^{\delta_{1}} \\
              \wedge &\ 1_{G_{1}} = l_{1}^{x_{i}}\cdot v_{1}^{\delta_{1}} \\
              \wedge &\ \frac{\hat{e}(g_{1},g_{2})}{\hat{e}(l_{3},\omega)} =\hat{e}(u,\omega)^{\!-\alpha\!-\beta}\!\cdot\!\hat{e}(l_{3},g_{2})^{x_{i}}\!\cdot\!\hat{e}(u,g_{2})^{\!-\delta_{1}\!-\delta_{1}}\! \cdot\!\hat{e}(h,g_{2})^{y_{i}}  \\
              \wedge &\ l_{4} = u_{0}^{x_{i}} \\
\end{array}
\right \}(m)$
\end{center}

\hspace{2.5mm}The linkable group signature on $m$ of user $i$ in an event $event$ is
$\sigma=$

\hspace{2.5mm}$(l_{1},l_{2},l_{3},l_{4},\Pi)$. Among $l_{4}$ is the tag for linking.

\begin{itemize}
\item \emph{\textbf{$Verify$}:} Known a group public key $GPK$, the signature $\sigma$ on $m$ for the specified $amount\in\{0,1\}^{\mathcal{*}}$, all members in the group can check the validity of the signature:

1. Compute $u_{0}=H_{0}(amount)$, then generates
\begin{center}
${\tilde a_{1}=v_{1}^{z_{\alpha}}\cdot l_{1}^{c}}$, \,\,\, ${\tilde a_{2}=v_{2}^{z_{\beta}}\cdot l_{2}^{c}}$, \,\,\,
\end{center}
\hspace{0.5cm} ${\tilde a_{3}=\hat{e}(u,\omega)^{\!-z_{\alpha}\!-z_{\beta}}\!\cdot\!\hat{e}(l_{3},g_{2})^{z_{x_{i}}}\!\cdot\!\hat{e}(u,g_{2})^{\!-z_{\delta_{1}}\!-z_{\delta_{1}}}\! \cdot\!\hat{e}(h,g_{2})^{z_{y_{i}}}\cdot \Big(\frac{\hat{e}(g_{1},g_{2})}{\hat{e}(l_{3},\omega)}\Big) ^{c}   \\}$ \,\,\,
\begin{center}
${\tilde a_{4}=l_{1}^{z_{x_{i}}}}\cdot v_{1}^{z_{\alpha}}$, \,\,\, ${\tilde a_{5}=l_{2}^{z_{x_{i}}}}\cdot v_{2}^{z_{\beta}}$, \,\,\, ${\tilde a_{6}=l_{4}^{c}}\cdot u_{0}^{z_{x_{i}}}$
\end{center}

2. On the basis of aforementioned value, and computes
 $$\tilde{c}=H_{1}(m,l_{1},l_{2},l_{3},l_{4},\tilde{a_{1}},\tilde{a_{2}},\tilde{a_{3}},\tilde{a_{4}},\tilde{a_{5}},\tilde{a_{6}})$$

3. Verify whether the equation $c=\tilde{c}$. It outputs 1 if equation is true and 0 otherwise.

\item \emph{\textbf{$Link$}:} Given different signature $(m,\sigma)$ and $(m',\sigma')$, anyone can publicly link if two signatures are signed by the identical signer for the same $amount$. Firstly verify if $\sigma$ and $\sigma'$ is valid for $m$ and $m'$ with $Verify$. Then it can be directly known $l_{4}$ from $\sigma$, $l_{4}'$ from $\sigma'$, decides whether $l_{4}=l_{4}'$.

\item \emph{\textbf{$Trace$}:} Given $(m,\sigma)$, If $\sigma$ is valid, SA tracing the original user by $A_{i}=\frac{l_{3}}{l_{1}^{k_{1}}\cdot l_{2}^{k_{1}}}$ with his private key $SK_{S}=(k_{1},k_{2})$ . Then SA gets the index $i$ in a group through the registration table given by the RA.


\end{itemize}

\subsection{Instantiation of the SoK}\label{subsec:sok}
The non-interactive zero-knowledge proof-of-knowledge $\Pi$ mentioned in subsection 4.2 actually is a signature of knowledge for message $m$. Here more details are given below.

Signer chooses $r_{\alpha},r_{\beta},r_{x},r_{y},r_{\delta_{1}},r_{\delta_{2}}$ at random from $z_{p}$, and computes

\begin{center}
$a_{1}=v_{1}^{r_{\alpha}}$,\ \ \ $a_{2}=v_{2}^{r_{\beta}}$,\ \ \
\end{center}
\begin{center}
$a_{3}=\hat{e}(u,\omega)^{\!-r_{\alpha}\!-r_{\beta}}\!\cdot\!\hat{e}(l_{3},g_{2})^{r_{x}}\!\cdot\!\hat{e}(u,g_{2})^{-r_{\delta_{1}}-r_{\delta_{2}}}\!\cdot\!\hat{e}(h,g_{2})^{r_{y}}$
\end{center}
\begin{center}
$a_{4}=l_{1}^{r_{x}}\cdot v_{1}^{-r_{\delta_{1}}}$,\ \ \ $a_{5}=l_{1}^{r_{x}}\cdot v_{2}^{-r_{\delta_{2}}}$,\ \ \ $a_{6}=u_{0}^{r_{x}}$
\end{center}

Then he sets $c=H_{1}(m,l_{1},l_{2},l_{3},l_{4},a_{1},a_{2},a_{3},a_{4},a_{5},a_{6})\in z_{p}$.

Subsequently he computes

\begin{center}
$z_{\alpha}=r_{\alpha}-c\alpha$,\ \ \ \ \ $z_{\beta}=r_{\beta}-c\beta$,

$z_{x}=r_{x}-c x$,\ \ \ \ \ $z_{y}=r_{y}-c y$,

$z_{\delta_{1}}=r_{\delta_{1}}-c\delta_{1}$,\ \ \ \ \ $z_{\delta_{2}}=r_{\delta_{2}}-c\delta_{2}$
\end{center}

Finally it outputs $\Pi$ is parsed as $(c,z_{\alpha},z_{\beta},z_{x},z_{y},z_{\delta_{1}},z_{\delta_{2}})$

Verifier computes the following six equation to verify $\Pi$:
$$v_{1}^{z_{\alpha}}=a_{1}\cdot l_{1}^{c}$$
$$v_{2}^{z_{\beta}}=a_{2}\cdot l_{2}^{c}$$
$$\hat{e}(u,\omega)^{\!-z_{\alpha}\!-z_{\beta}}\cdot\hat{e}(l_{3},g_{2})^{z_{x}}\cdot \hat{e}(u,g_{2})^{\!-z_{\delta_{1}}\!-z_{\delta_{1}}} \cdot \hat{e}(h,g_{2})^{z_{y}}=a_{3}\cdot\bigg(\frac{\hat{e}(g_{1},g_{2})}{\hat{e}(l_{3},\omega)}\bigg)^{c}$$
$$l_{1}^{z_{x}}\cdot v_{1}^{-z_{\alpha}}=a_{4}$$
$$l_{1}^{z_{x}}\cdot v_{2}^{-z_{\delta_{2}}}=a_{5}$$
$$u_{0}^{z_{x}}=a_{6}\cdot l_{4}^{c} $$

Verifier outputs 1 if $c=H_{1}(m,l_{1},l_{2},l_{3},l_{4},a_{1},a_{2},a_{3},a_{4},a_{5},a_{6})\in Z_{p}$, otherwise 0.

\subsection{Correctness}
The signature $\sigma$ produced in section \ref{subsec:protocol} is a signature of knowledge with respect to message $m$, which is the transcript of zero proof of knowledge about a pair $(A_{i},x_{i},y_{i})$ satisfies $A_{i}^{\gamma+x_{i}}h^{y_{i}}=g_{1}$. Therefore, verifier will accept the signature if the transcript is verified in the light of the way in section \ref{subsec:sok} correctly.

In addition, the first three elements of any signature $\sigma$ contains $(l_{1},l_{2},l_{3})=(v_{1}^{\alpha},v_{2}^{\beta},A_{i}\cdot u^{\alpha+\beta})$, which is the Linear encryption of $A_{i}$. Supervision authority, who owns $SK_{S}=(k_{1},k_{2})$, can decrypt it correctly and recover index $i$ corresponding $A_{i}$ in a group of $n$ members.

The same tag will be directly linked through comparing $l_{4}$, which is the fourth components of the signature $\sigma$. It means that the signature is generated by the same signer for the same $amount$.

\subsection{Extensions}
Our registration protocol$(Join)$ can achieve a stronger level of security to protect the privacy of user's identity, which is statistically zero-knowledge\cite{ateniese}. Specially, at the beginning of protocol execution, user randomly selects $\tilde{y_{i}},\tilde{r_{i}}\in_{R}\,Z_{p}$, sends $C_{1}=g_{1}^{\tilde{y_{i}}}g_{2}^{\tilde{r_{i}}}$ to the RA, in the meantime user executes the knowledge of the representation of $C_{1}$ to bases $g_{1}$ and $g_{2}$. Then if RA agree the proof, he selects $\alpha_{i},\beta_{i}\in_{R}\,Z_{p}$ at random, and sends $(\alpha_{i},\beta_{i})$ to user. he computes $y_{i}=\alpha_{i}\tilde{y_{i}}+\beta_{i}$ and sends $C_{2}=h^{y_{i}}$ to RA.

\section{Security Analysis}\label{sec5}
In this section, we analyze the security of the linkable group signature construction in the random oracle mode. The security of our scheme is guaranteed by following lemmas.

\textbf{Lemma 1.} \emph{Our scheme satisfies CPA-full-anonymous if Linear encryption against chosen-plaintext attacks on $G_{1}$ in the random oracle model.}

\textbf{Proof.} We present it by reducing. Suppose there exists an adversary $\mathcal{A}$ that breaks the anonymity of the linkable group signature, then we can construct a algorithm $\mathcal{A'}$ can break CPA-security of Linear encryption on $G_{1}$. As described in \cite{boneh}, Linear encryption can against a chosen-plaintext attack, it implies that adversary $\mathcal{A}$ wins full-anonymous game with negligible advantage.

$\mathcal{A'}$ is given a tuple $(v_{1},v_{2},u)$ as public key of Linear encryption, then it runs $Setup$ algorithm to generate the group public parameter $GPK$ and sends it to $\mathcal{A}$.

\textbf{Join Queries:} $\mathcal{A}$ can randomly choose $y_{i}\in Z_{p}$ and query the random oracle $\mathcal{O_{J}}$, $\mathcal{A'}$ executes $Join$ protocol and responds with $Cert=(A_{i},x_{i})$.

\textbf{Hash Queries:} $\mathcal{A}$ is given $u_{0}$ randomly selects from $G_{1}$ if he queries $H_{0}$ with $amount$, when he requests $H_{1}$, elements randomly chose from $Z_{p}$ is responded.

\textbf{Challenge:} $\mathcal{A}$ randomly picks two users corresponding their public keys $Y_{i_{1}}^{\ast}=h^{y_{i_{1}}}$ and $Y_{i_{2}}^{\ast}=h^{y_{i_{2}}}$, message $m^{\ast}$ and amount $amount^{\ast}$, $\mathcal{A'}$ provides $A_{i_{1}}$ and $A_{i_{2}}$ as challenge message, and requests to the challenger in indistinguishability game of Linear encryption. The challenger responds with $(l_{1},l_{2},l_{3})$, which is the ciphertext of $A_{i_{b}}$ such that $b\in\{0,1\}$. $\mathcal{A'}$ also chooses a bit $b\in\{0,1\}$ randomly, and sets $l_{4}=u_{0}^{x_{i_{b}}}$, where $u_{0}=H_{0}(amount^{\ast})$. Then $\mathcal{A'}$ gets a tuple transcript $(l_{1},l_{2},l_{3},l_{4},a_{1},a_{2},a_{3},a_{4},a_{5},a_{6},c,z_{\alpha},z_{\beta},z_{x},z_{y},z_{\delta_{1}},z_{\delta_{2}})$
by calling simulator even $\mathcal{A'}$ doesn't know real $\alpha,\beta,x_{i},y_{i}$, which is indistinguishing with real proof of zero-knowledge protocol about $(A_{i_{b}},x_{i_{b}},y_{i_{b}})$. In addition, algorithm $\mathcal{A'}$ has to make $H_{1}(m,l_{1},l_{2},l_{3},l_{4},a_{1},a_{2},a_{3},a_{4},a_{5},a_{6})=c$ is true, it stops if there has a collision, otherwise it returns $\sigma^{\ast}=(l_{1},l_{2},l_{3},l_{4},c,z_{\alpha},z_{\beta},z_{x},z_{y},z_{\delta_{1}},z_{\delta_{2}})$ to $\mathcal{A}$ as challenge signature.

\textbf{Guess:} $\mathcal{A}$ outputs a guess $b'$, meanwhile $\mathcal{A'}$ sends $b'$ as solution to the challenger and it wins when $b=b'$. $\mathcal{A}$ wins in anonymity games when $\mathcal{A'}$ succeeds in indistinguishable game of Linear encryption respect to $A_{i_{b}}$. If the advantage of $\mathcal{A}$ is regarded $Adv_{\mathcal{A}}^{CPA-anoy}=\epsilon$, then algorithm $\mathcal{A'}$ against a chosen-plaintext attack of the Linear encryption with probability $1/2+\epsilon$.

\textbf{Lemma 2.} \emph{Our scheme satisfies full-traceability if VR-SDH assumption holds on $(G_{1},G_{2})$.}

\textbf{Proof.} We present it by reducing. We borrowed the skill with respect to the security of full-traceability in \cite{boneh}, which is divided into three phases. Firstly, a framework invokes a full-traceability game interact with an adversary $\mathcal{A}$ is given. Secondly, instantiating the framework for different types of adversaries. Thirdly, computing an VR-SDH solutions by applying Forking Lemma \cite{boneh} to the instantiation.

\textbf{Phase I.} Suppose there exists an adversary $\mathcal{A}$ that breaks the full-traceability of the linkable group signature, then it can be constructed a framework interact with $\mathcal{A}$ as follows.

Framework is given group public parameter $GPK$ and a sets of VR-SDH tuples $(A_{i},x_{i},y_{i})$ for $i=1,...,n$ most of them satisfied $\hat{e} \ (A_{i},\omega{g_{2}^{x_{i}}})=\hat{e} \ (g_{1}{h^{-y_{i}},g_{2}})$, otherwise for $i=\diamond$, it means that $x_{i}$ is unknown, which corresponding to $(A_{i},y_{i})$. Then it sends $GPK$ and the private key of SA that $SK_{S}=(k_{1},k_{2})$ to $\mathcal{A}$.

\textbf{Hash Queries:} $\mathcal{A}$ is given $u_{0}$ randomly selects from $G_{1}$ if he queries $H_{0}$ with $amount$, when he requests the $H_{1}$ of $(m,l_{1},l_{2},l_{3},l_{4},a_{1},a_{2},a_{3},a_{4},a_{5},a_{6})$, framework randomly chooses $c\in Z_{p}$, then returns $c$ to $\mathcal{A}$.

\textbf{Join Queries:} $\mathcal{A}$ can randomly chooses $y_{i}\in Z_{p}$ and query the random oracle $\mathcal{O_{J}}$, when $i\neq\diamond$ framework returns $Cert=(A_{i},x_{i})$ to $\mathcal{A}$. Otherwise, it terminates.

\textbf{Sign Queries:} $\mathcal{A}$ asks for a signature of $ith$ users on amount $amount$ and message $m$. If $i\neq \diamond$, framework generates a signature $\sigma$ with real private key $(A_{i},x_{i},y_{i})$. If $i=\diamond$, it computes $u_{0}=H_{0}(amount)$ and sets $l_{1},l_{2},l_{3},l_{4}$ to be $v_{1}^{\alpha}, v_{2}^{\beta}, A_{i}\cdot u ^{\alpha+\beta}, u_{0}^{x_{i}}$ for some random $\alpha,\beta\in Z_{p}$. Then it obtained a tuple transcript $(l_{1},l_{2},l_{3},l_{4},a_{1},a_{2},a_{3},a_{4},a_{5},a_{6},c,z_{\alpha},z_{\beta},z_{x},z_{y},z_{\delta_{1}},z_{\delta_{2}})$ by calling simulator, where $\sigma=(l_{1},l_{2},l_{3},l_{4},c,z_{\alpha},z_{\beta},z_{x},z_{y},z_{\delta_{1}},z_{\delta_{2}})$ is originated. Moreover, it must mend the hash value in $H_{1}(m,l_{1},l_{2},l_{3},l_{4},a_{1},a_{2},a_{3},a_{4},a_{5},a_{6})$ be $c$, it terminates if a collision is occurred. Then it returns $\sigma$ to $\mathcal{A}$.

\textbf{Forge.} $\mathcal{A}$ outputs a signature $\sigma=(l_{1},l_{2},l_{3},l_{4},c,z_{\alpha},z_{\beta},z_{x},z_{y},z_{\delta_{1}},z_{\delta_{2}})$ on message $m$, which can be traced to obtain $\tilde{A}$ with $(k_{1},k_{2})$. If $\tilde{A}\neq A_{i}$ $(i=1,2,...,n)$, framework outputs $\sigma$. If $\tilde{A}= A_{i^{\ast}}$, $x_{i^{\ast}}=\diamond$ and $Y_{i^{\ast}}$ is not queried to join oracle occurs currently, it outputs $\sigma$, Otherwise, if $x_{i^{\ast}}\neq \diamond$, it exists.

\textbf{Phase II.} Due to different cases, we instantiate them by two types of forgers. Forger $\mathcal{A}_{1}$ is given $(g_{1},g_{2},\omega)$ and $n$ SDH-VR pairs $(A_{i},x_{i},y_{i})$, framework interacts with $\mathcal{A}_{1}$ according to aforementioned process. If it can be perfectly simulated, based on the success of $\mathcal{A}_{1}$, the framework succeeds. In this case, $\mathcal{A}_{1}$ succeeds with probability $\epsilon$.

Forger $\mathcal{A}_{2}$ is given $(g_{1},g_{2},\omega)$ and $n-1$ VR-SDH pairs $(A_{i},x_{i},y_{i})$, then framework randomly selects $A_{i^{\ast}}$ from $G_{1}$, and sets $x_{i^{\ast}}=\diamond$. Let these pairs constitute a group with $n$ members. If $\mathcal{A}_{2}$ forges a valid signature $\sigma^{\ast}$ that can be traced to $A_{i^{\ast}}$, at the same time, not for a moment does $\mathcal{A}_{2}$ query to $\mathcal{O}_{J}$ at $y_{i^{\ast}}$, then the framework proclaims success. so that $\mathcal{A}_{2}$ outputs a imitative linkable group signature that can trace to the user of $i^{\ast}$ with probability $\epsilon/n$.

\textbf{Phase III.} We can obtain the solution of VR-SDH problem by applying Forking Lemma to different forgers \cite{boneh}. Linkable group signature can be indicated as $(m,\sigma_{0},c,\sigma_{1})$, where $\sigma_{0}=(l_{1},l_{2},l_{3},l_{4},a_{1},a_{2},a_{3},a_{4},a_{5},a_{6})$, $\sigma_{1}=(z_{\alpha},z_{\beta},z_{x},z_{y},z_{\delta_{1}},z_{\delta_{2}})$. It can be concluded that the framework obtains a set of forgery $(m,\sigma_{0}',c',\sigma_{1}')$ with probability $(\epsilon-1/p)^{2}/4$ when $\mathcal{A}_{1}$ succeeds, or $(\epsilon/n-1/p)^{2}/4$ when $\mathcal{A}_{2}$ succeeds. Furthermore, it
generates another set of forgery $(m,\sigma_{0}'',c'',\sigma_{1}'')$ with probability $(\epsilon-1/p)^{2}/4q_{H}$ if $\mathcal{A}_{1}$ succeeds, or $(\epsilon/n-1/p)^{2}/4q_{H}$ if $\mathcal{A}_{2}$ succeeds, where $q_{H}$ is the number of hash function queries.

There exists a extractor can extract a solution of VR-SDH problem for two forgeries $(m,\sigma_{0}',c',\sigma_{1}')$ and $(m,\sigma_{0}'',c'',\sigma_{1}'')$, the framework declares success when extracted tuple $(A,x,y)$ is not distributed in those whose $x$ is known.

On the basis of aforesaid, if interacts with forger $\mathcal{A}_{1}$, framework can solve VR-SDH problem with probability $(\epsilon-1/p)^{2}/16q_{H}$. Otherwise interacts with forger $\mathcal{A}_{2}$, it can return $(A,x,y)$ as the solution of VR-SDH problem with $(\epsilon/n-1/p)^{2}/16q_{H}$. Due to it is widely believed the VR-SDH problem is intractable, so forgers whether $\mathcal{A}_{1}$ or $\mathcal{A}_{2}$ succeeds to forge a linkable group signature with negligible advantage.

\textbf{Lemma 3.} \emph{Our scheme satisfies linkability if VR-SDH assumption holds on $(G_{1},G_{2})$.}

\textbf{Proof.} We present it by reducing. If the signature can be linked, it must be signed by identical signer for the same $amount$ with uniform private key $(A_{i},x_{i},y_{i})$. Suppose there exists an PPT adversary $\mathcal{A}$ that breaks the linkability of the linkable group signature, then it must be constructed an another PPT algorithm $\mathcal{A'}$ can solve VR-SDH problem with non-negligible probability.

\textbf{Join Queries:} $\mathcal{A}$ can randomly choose $y_{i}\in Z_{p}$ and query the random oracle $\mathcal{O_{J}}$, $\mathcal{A'}$ executes $Join$ protocol and responds with $Cert=(A_{i},x_{i})$.

\textbf{Hash Queries:} $\mathcal{A}$ is given $u_{0}$ randomly selects from $G_{1}$ if he queries $H_{0}$ with $amount$, when he requests $H_{1}$, elements randomly chose from $Z_{p}$ is responded.

\textbf{Sign Queries:} $\mathcal{A}$ asks for a signature of users private key $(A_{i},x_{i},y_{i})$ for amount $amount$ on message $m$. It is given a signature $\sigma$ by the challenger, who performs $Sign$ algorithm.

\textbf{Forge.} $\mathcal{A}$ outputs $(m_{i_{1}}',\sigma_{i_{1}}')$ and $(m_{i_{2}}',\sigma_{i_{2}}')$ with respect to $(A_{i}',x_{i}',y_{i}')$, where $(A_{i}',x_{i}',y_{i}')$ is not queried to the signature oracle $\mathcal{O_{S}}$. $\mathcal{A}$ wins if undermentioned cases occurs concurrently.
\begin{center}
$Verify(GPK,m_{i_{1}}',\sigma_{i_{1}}')=1,$
$Verify(GPK,m_{i_{2}}',\sigma_{i_{2}}')=1,$
$l_{4}^{i_{1}'}\neq l_{4}^{i_{2}'}.$
\end{center}
where $l_{4}^{i_{1}'}$ and $l_{4}^{i_{2}'}$ are the fourth components of $\sigma_{i_{1}}'$ and $\sigma_{i_{2}}'$. Then there can be constructed a PPT algorithm $\mathcal{A'}$ settles VR-SDH problem by computing $(A_{i}'',x_{i}'',y_{i}'')$ other than $(A_{i}',x_{i}',y_{i}')$ with non-negligible probability. Due to it is widely believed VR-SDH is difficult, therefore $\mathcal{A}$ succeeds to break the linkability of linkable group
signature with negligible probability.

\section{Performance}\label{sec6}
\subsection{Implementation analysis.}
In this section, we discuss the implementation analysis of our scheme. We conduct the simulations 
on a Win 10 64-bit desktop with 8.00 GB RAM and Inter(R) Core(TM) i5-7400 CPU @ 3.00 GHz. All the algorithms are written in C++ language and invoke the Miracl library for elliptic curve cryptography. We use Visual Studio 2012 to compile all the programs. There are six algorithms named $Setup$, $Join$, $Sign$, $Verify$, $Link$, $Trace$, we execute them under the number of individuals varied from 3 to 10 and test each algorithm 20 times separately on the desktop. Then we calculated the average running time of each algorithm for different size of group as shown in Figure \ref{fig2}.

\begin{figure}[htbp]
  \centering
  \includegraphics[width=0.8\textwidth]{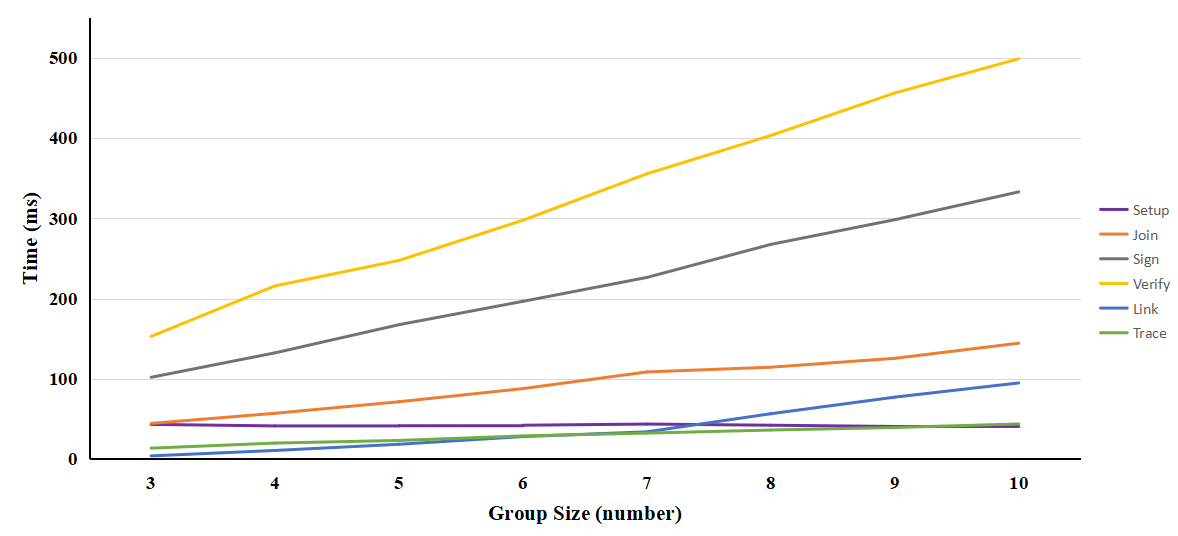}\\
  \caption{Time cost of LGS algorithms for different size group}\label{fig2}
\end{figure}

As displayed in Figure \ref{fig2}, group size is determined by the number of group members. The time cost of $Setup$ algorithm tend to be a constant with the increasing of group members, nearly 41.125ms. This is due to initialization variables are fixed per routs, hence, it takes approximately the identical time. Moreover, the time consumption of $Join$ algorithm grows linearly as member grows in a group. The size of group varies from 3 to 10 and $Join$ algorithm takes 44.25ms at least and 144.286ms at most, this result is rational. The implementation time that $Sign$ algorithm and $Verify$ algorithm costs respectively are exceedingly fast since the $Sign$ algorithm chooses several random values meanwhile executes some exponentiation and pairing operations, it expends 196.5ms when group size is 6, while the $Verify$ algorithm raises more quicker than $Sign$, it costs 297.667ms if the size of group is 6. The results are consistent with our empirical analysis due to it needs to perform more exponentiation and pairing for a generated signature. With the increasing number of group members from three to ten, the time cost concerning $Link$ algorithm also increases but still tiny where the largest is 94.714ms when group size is 10 because it just calculates a comparison. Similarly, the $Trace$ algorithm grows linearly and the increments taper off, it expenses about 41.134ms and 43.574ms respectively when there are 9 members and 10 members in a group. All statements aforementioned are consistent with our empirical analysis.

\section {Conclusion and future work}\label{sec7}

Cryptocurrencies have gained increasing recognition. Furthermore, The regulation is also indispensable in order to prevent the abuse of cryptocurrencies. In this paper, we proposed a fresh linkable group signature based on the Consortium Blockchain to achieve the goal which tracing the real-world identity in anonymous cryptocurrencies. Then we proved our scheme satisfied the desirable security properties of linkable group signature. At last, the implementations testify the feasibility of our scheme.






\end{document}